\documentclass[structabstract, longauth]{aa}
\pdfoutput=1
\usepackage{graphicx}
\usepackage{txfonts}
\usepackage{natbib}
\begin{document}
\title{Study of the $\gamma$-ray source 1AGL~J2022+4032 in the Cygnus Region}

\author{
A.~W.~Chen\inst{3} \and G.~Piano\inst{1,2,11} \and
M.~Tavani\inst{1,2} \and A.~Trois\inst{1} \and
G. Dubner\inst{20} \and E.~Giacani\inst{20} \and
A.~Argan\inst{1} \and
G.~Barbiellini\inst{6} \and
A. Bulgarelli\inst{5} \and
P.~Caraveo\inst{3} \and
P.~W.~Cattaneo\inst{7} \and 
E.~Costa\inst{1} \and
F.~D'Ammando\inst{1,2} \and
G.~De~Paris\inst{1} \and 
E. Del Monte\inst{1} \and
G.~Di~Cocco\inst{5} \and
I.~Donnarumma\inst{1} \and 
Y.~Evangelista\inst{1} \and
M.~Feroci\inst{1} \and 
A.~Ferrari\inst{4,18} \and
M.~Fiorini\inst{3} \and
F.~Fuschino\inst{5} \and 
M.~Galli\inst{8} \and
F.~Gianotti\inst{5} \and  
A.~Giuliani\inst{3} \and
M.~Giusti\inst{1,3} \and 
C.~Labanti\inst{5} \and
F.~Lazzarotto\inst{1} \and 
P.~Lipari\inst{9} \and
F.~Longo\inst{6} \and 
M.~Marisaldi\inst{5} \and
S.~Mereghetti\inst{3} \and
E.~Moretti\inst{6} \and 
A.~Morselli\inst{11} \and
L.~Pacciani\inst{1} \and
A.~Pellizzoni\inst{19} \and
F.~Perotti\inst{3} \and 
P.~Picozza\inst{2,11} \and
M.~Pilia\inst{12,19} \and
M.~Prest\inst{12} \and 
G.~Pucella\inst{13} \and
M.~Rapisarda\inst{13} \and 
A.~Rappoldi\inst{7} \and
S. Sabatini \inst{1,11} \and 
E.~Scalise\inst{1} \and
P.~Soffitta\inst{1} \and 
E. Striani\inst{2,11} \and
M.~Trifoglio\inst{5} \and
E.~Vallazza\inst{6} \and
S.~Vercellone\inst{17} \and
V.~Vittorini\inst{1,2} \and
A.~Zambra\inst{3} \and
D.~Zanello\inst{9} \and 
C.~Pittori\inst{14} \and
P.~Giommi\inst{14} \and
F.~Verrecchia\inst{14} \and
F.~Lucarelli\inst{14} \and
P.~Santolamazza\inst{14} \and
S.~Colafrancesco\inst{14} \and
L.A.~Antonelli\inst{19} \and 
L.~Salotti\inst{15} }

\institute {INAF/IASF-Roma, I-00133 Roma, Italy
\and Dip. di Fisica, Univ. Tor Vergata, I-00133 Roma, Italy 
\and INAF/IASF-Milano, I-20133 Milano, Italy
\and CIFS-Torino, I-10133 Torino, Italy
\and INAF/IASF-Bologna, I-40129 Bologna, Italy
\and Dip. Fisica and INFN Trieste, I-34127 Trieste, Italy 
\and INFN-Pavia, I-27100 Pavia, Italy
\and ENEA-Bologna, I-40129 Bologna, Italy
\and INFN-Roma La Sapienza, I-00185 Roma, Italy
\and CNR-IMIP, Roma, Italy
\and INFN-Roma Tor Vergata, I-00133 Roma, Italy
\and Dip. di Fisica, Univ. Dell'Insubria, I-22100 Como, Italy
\and ENEA Frascati,  I-00044 Frascati (Roma), Italy
\and ASI Science Data Center, I-00044 Frascati(Roma), Italy
\and Agenzia Spaziale Italiana, I-00198 Roma, Italy 
\and INAF-Osservatorio Astron. di Roma, Monte Porzio Catone, Italy
\and  INAF-IASF-Palermo, via U. La Malfa 15, I-90146 Palermo, Italy
\and Dip. Fisica, Universit\'a di Torino, Turin, Italy
\and INAF-Osservatorio Astronomico di Cagliari, 
	localita' Poggio dei Pini, strada 54, I-09012 Capoterra, Italy  
\and Institute of Astronomy and Space Physics, University of Buenos Aires, 
Ciudad Universitaria, (1428) Ciudad de Buenos Aires, Argentina 
}

   \date{Received 24 June 2010; accepted 20 September 2010}

\abstract
{Identification of $\gamma$-ray-emitting Galactic sources is a long-standing 
problem in astrophysics. One such source, \object{1AGL J2022+4032}, coincident 
with the interior of the radio shell of the supernova remnant 
Gamma Cygni (\object{SNR G78.2+2.1}) in the Cygnus Region, has recently been 
identified by \textit{Fermi} as a $\gamma$-ray pulsar, 
LAT~\object{PSR J2021+4026}.}
{We present long-term observations of 1AGL~J2022+4032 with the 
\textit{AGILE} $\gamma$-ray telescope, measuring its flux and light curve.}
{We compare the light curve of 1AGL~J2022+4032 with that of 
\object{1AGL J2021+3652} (\object{PSR J2021+3651}), showing that the flux 
variability of 1AGL J2022+4032 appears to be greater than the level predicted 
from statistical and systematic effects and producing detailed simulations
to estimate the probability of the apparent observed variability.}
{We evaluate the possibility that the $\gamma$-ray emission may be due to the
superposition of two or more point sources, some of which may be
variable, considering a number of possible counterparts.}
{We consider the possibility of a nearby X-ray quiet microquasar contributing 
to the flux of 1AGL~J2022+4032 to be more likely than the hypotheses of a 
background blazar or intrinsic $\gamma$-ray variabilty of LAT~PSR J2021+4026.}


\keywords{Gamma rays: stars:
	\object{1AGL J2022+4032}, \object{1FGL J2021.5+4026}, 
	\object{3EG J2020+4017} ---
	pulsars: LAT~\object{PSR J2021+4026}
	}

\authorrunning{A. W. Chen et al.}
\titlerunning{Study of the $\gamma$-ray source 1AGL~J2022+4032 in the 
	Cygnus Region}

   \maketitle
%

\section{Introduction}

Identification of the Galactic sources emitting $\gamma$-rays with energies 
above 100 MeV is a long-standing problem in astrophysics.  With the launch of 
the \textit{AGILE} $\gamma$-ray telescope in 2007 \citep{agile} and 
\textit{Fermi} in 2008 \citep{fermi}, a great deal of progress has been made. 
In particular, \textit{Fermi} and \textit{AGILE} have shown that the vast 
majority of Galactic $\gamma$-ray sources are probably pulsars and pulsar wind 
nebulae \citep{fermibsl}.  However, new source classes have 
also been identified, including microquasars such as 
\object{Cygnus X-1} \citep{cygx1}, 
\object{Cygnus X-3} \citep{agilecygx3,fermicygx3}, 
\object{LS5039} \citep{fermils5039} and 
LSI$+61^{\circ}303$ (\object{LSI+61 303}) \citep{fermilsi}, and Wolf-Rayet 
stars such as \object{Eta Carinae} \citep{etacar}.

As the identification of the remaining unidentified \textit{EGRET} sources and 
the newly found \textit{AGILE} and \textit{Fermi-LAT} sources continues, the 
problem of source confusion will become increasingly serious.  Although the 
angular resolution of both \textit{AGILE} and \textit{Fermi} are higher 
than that of \textit{EGRET}, they also are sensitive to much lower fluxes, 
\textit{Fermi-LAT} in particular.  
As a result, $\gamma$-ray error contours will not only continue to contain 
many plausible source counterparts, but will be increasingly likely to contain 
multiple real $\gamma$-ray sources above the nominal flux threshold.

\begin{figure}[htbp]
   \centering
	\resizebox{\hsize}{!}{
		\includegraphics{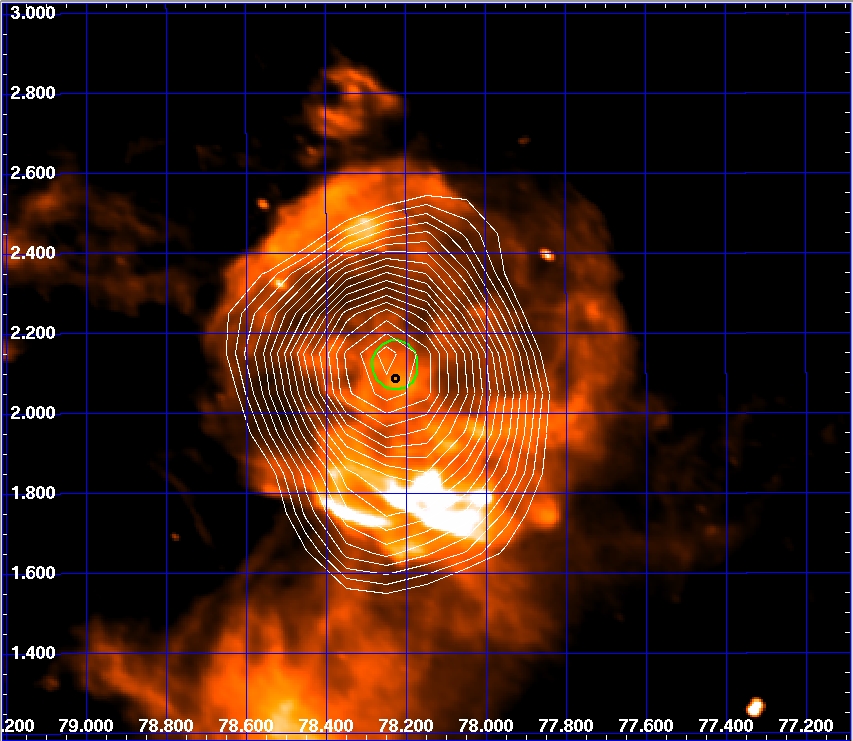}
	}
   \caption{SNR Gamma Cygni (G78.2+2.1) in galactic coordinates. 
	DRAO Radio telescope, wavelength=21.1 cm. \textit{White} contour levels: 
	\textit{AGILE-GRID} intensity contour levels - related to 
	Figure \ref{Cygnus-region} - (pixel size $0.1^{\circ}$), starting from 
	0.00085 in steps of 0.00002 (intensity per pixel); \textit{green} 
	contour: \textit{AGILE-GRID} 95\% confidence level for 
	$E \ge 100~$MeV; \textit{black} circle: LAT~PSR J2021+4026.} 
   \label{gamma_cyg_snr}
\end{figure}

In this paper we provide evidence that \object{1AGL J2022+4032} in the Cygnus 
region \footnote{Throughout this paper we refer to the Cygnus region, although 
physically this direction lies along the tangents of at least three different 
Galactic spiral arms at different distances, from the local 
$\approx1$~kpc Orion spur, to the Perseus arm, to the Outer arm at 
$>8$~kpc.} \citep{agilecat} may be one of these sources.  We present a study 
of the time variability of the $\gamma$-ray flux of 1AGL~J2022+4032 
based on long term observations of the source by \textit{AGILE}. This source 
was first discovered by COS-B as \object{2CG078+2}, and was listed as the 
unidentified source \object{3EG J2020+4017} in the 3rd \textit{EGRET} catalog 
\citep{egretcat}. It is located (Figure~\ref{gamma_cyg_snr}) within the wide 
radio shell of the prominent SNR Gamma Cygni (\object{SNR G78.2+2.1}). 
Searches for X-ray counterparts of the source were performed with a number of 
X-ray telescopes, including {\it ROSAT} \citep{brazier}, {\it ASCA} 
\citep{uchiyama}, {\it INTEGRAL} \citep{bykov}, and {\it Chandra} 
\citep{becker, weisskopf}. 
Using a blind search technique, \textit{Fermi-LAT} was finally able to 
identify this source, \object{1FGL J2021.5+4026} \citep{fermibsl, 1FGL}, as 
$\gamma$-ray pulsar LAT~\object{PSR J2021+4026} \citep{fermipulsar}. The 
$\gamma$-ray flux and spectra are consistent with a pulsar origin.  However, 
$\gamma$-ray pulsars show no variability over time scales of weeks to months; 
indeed, significant variability on these time scales would be difficult to 
reconcile with theoretical models. 

Beginning in November 2007,  \textit{AGILE} data from this source have 
revealed signs of variable $\gamma$-ray emission, as reported in 
ATel \#1492 \citep{at1492}, \#1547 \citep{at1547} and 
\#1585 \citep{at1585}.  The evidence for $\gamma$-ray variability combined 
with its relatively high unpulsed $\gamma$-ray fraction could indicate either 
the presence of an additional $\gamma$-ray source coincident with 
1FGL~J2021.5+4026 or variability of the $\gamma$-ray pulsar LAT~PSR J2021+4026.


\section{Observations and Data Analysis}

\begin{figure*}[htbp]
   \centering
   \resizebox{\hsize}{!}{
	\includegraphics[width=12cm]{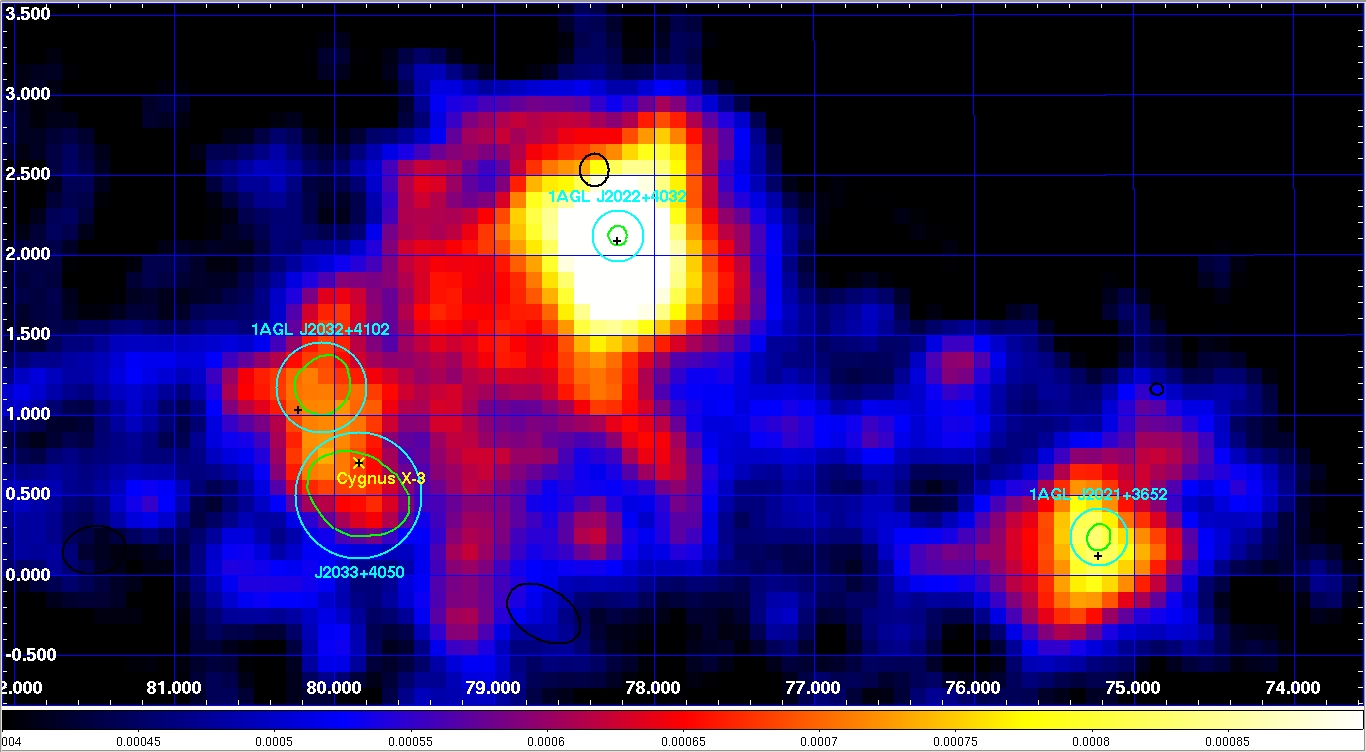}
	}
   \caption{Cygnus Region in galactic coordinates, $\gamma$-ray intensity 
	map for $E \ge 100~$MeV. Deep integration \textit{AGILE-GRID} data 
	(November 2007 - August 2009). Pixel size = $0.1^{\circ}$ with 3-pixel 
	Gaussian smoothing. \textit{Green} contours: \textit{AGILE-GRID} 
	95\% confidence level; \textit{Cyan} contours: \textit{AGILE-GRID} 
	statistical + systematic error $(0.1^{\circ})$; \textit{Black} 
	contours: \textit{Fermi-LAT} (1-year catalog), statistical error only, 
	crosses shown for contours too small to be visible. The \textit{Green} 
	contours have been calculated with a multi-source likelihood analysis, 
	using four persistent sources (Table \ref{sources}).}
   \label{Cygnus-region}
\end{figure*}

\begin{table*}[htbp]
   \centering
   \caption{$\gamma$-ray sources in the Cygnus Region.} \label{sources}
   \begin{tabular}{c c c c}
	\hline\hline
	Name  & Position  & $\sqrt{\mathrm{TS}}$  & Flux\tablefootmark{a}   \\
	\hline\hline
	1AGL J2021+3652, $E \ge 100~$MeV &   (l, b) = (75.22, 0.24) $\pm$ $0.08^{\circ}$ (stat) $\pm$ $0.10^{\circ}$ (syst) &         25.02        &   60  $\pm$   3   (stat)  $\pm$ 10\%  (syst) \\
	\hline
	1AGL J2022+4032, $E \ge 100~$MeV             &   (l, b) = (78.23, 2.12) $\pm$ $0.06^{\circ}$ (stat) $\pm$ $0.10^{\circ}$ (syst) &         39.64        &  131  $\pm$   4   (stat)  $\pm$ 10\%  (syst) \\
	\hline
	1AGL J2032+4102, $E \ge 100~$MeV   &   (l, b) = (80.08, 1.18) $\pm$ $0.18^{\circ}$ (stat) $\pm$ $0.10^{\circ}$ (syst) &         10.82        &   $37\pm 4$   (stat)  $\pm$ 10\%  (syst) \\
	\hline
	J2033+4050\tablefootmark{b}, $E \ge 100~$MeV      &   (l, b) = (79.84, 0.50) $\pm$ $0.29^{\circ}$ (stat) $\pm$ $0.10^{\circ}$ (syst) &          5.17        &   14  $\pm$   3   (stat)  $\pm$ 10\%  (syst) \\
	\hline\hline
	1AGL J2021+3652, $E \ge 400~$MeV             & (l, b) = (75.16, 0.23) $\pm$ $0.07^{\circ}$ (stat) $\pm$ $0.10^{\circ}$ (syst) &          23.07         &   17  $\pm$   1   (stat)  $\pm$ 10\%  (syst)   \\
	\hline
	1AGL J2022+4032, $E \ge 400~$MeV           & (l, b) = (78.21, 2.12) $\pm$ $0.05^{\circ}$ (stat) $\pm$ $0.10^{\circ}$ (syst) &          33.80         &   33  $\pm$   1   (stat)  $\pm$ 10\%  (syst)   \\
	\hline
	1AGL J2032+4102, $E \ge 400~$MeV & (l, b) = (80.05, 0.98) $\pm$ $0.14^{\circ}$ (stat) $\pm$ $0.10^{\circ}$ (syst) &          10.59         &    9  $\pm$   1   (stat)  $\pm$ 10\%  (syst)   \\
	\hline\hline
   \end{tabular}
   \\
   \tablefoot{
	\tablefoottext{a}{$\gamma$-ray fluxes in units of 
	$10^{-8}\mathrm{~photons~cm}^{-2}\mathrm{~s}^{-1}$.}
	\tablefoottext{b}{Positionally consistent with \object{Cygnus X-3}.}
   }
\end{table*}

\textit{AGILE} has been in orbit since April 2007 and has observed the Cygnus 
region numerous times, beginning in November 2007.  We used these observations 
to characterize the $\gamma$-ray variability of the sources in the Cygnus 
region using the following procedure. First, we used the software tool 
\verb+AG_multi2+, one of the \textit{AGILE} Scientific Tools, to perform 
multi-source likelihood analysis on the deep-integration \textit{AGILE-GRID} 
data using the \verb+FM3.119_2+ filter. This analysis revealed 
$\gamma$-ray emission from four point sources: \object{1AGL J2022+4032}, 
\object{1AGL J2021+3652}, \object{1AGL J2032+4102} and a persistent faint 
source consistent with the position of 
\object{Cygnus X-3} (Figure~\ref{Cygnus-region}).
The average fluxes and positions are shown in Table~\ref{sources}.
\textit{Fermi} source 1FGL~J2020.0+4049 was not detected in either energy 
range due to the combination of its proximity to 1AGL~J2022+4032 and its low 
flux.

\begin{figure*}[hbtp]
   \centering
   \resizebox{\hsize}{!}{
	\begin{tabular}{cc}
		\includegraphics{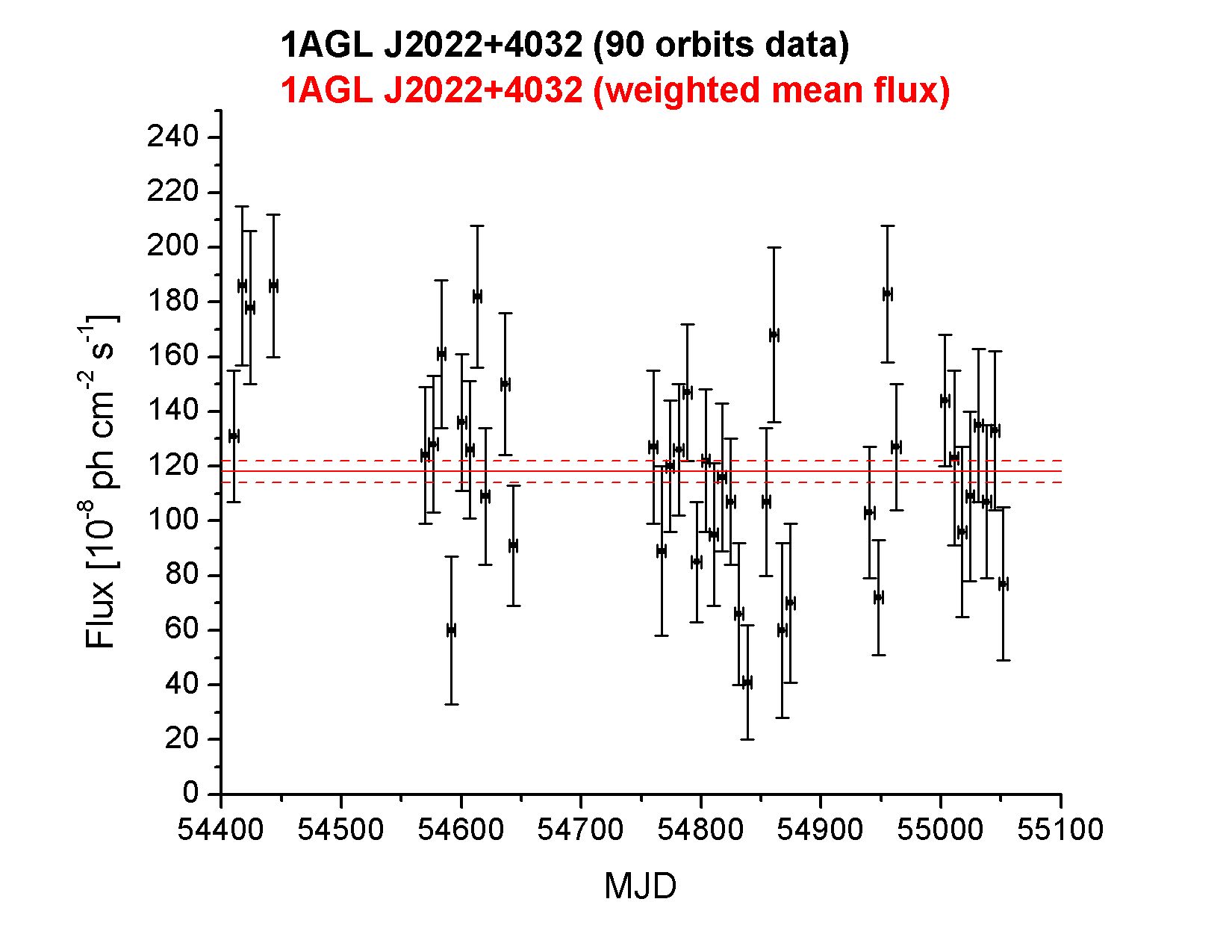} & 
		\includegraphics{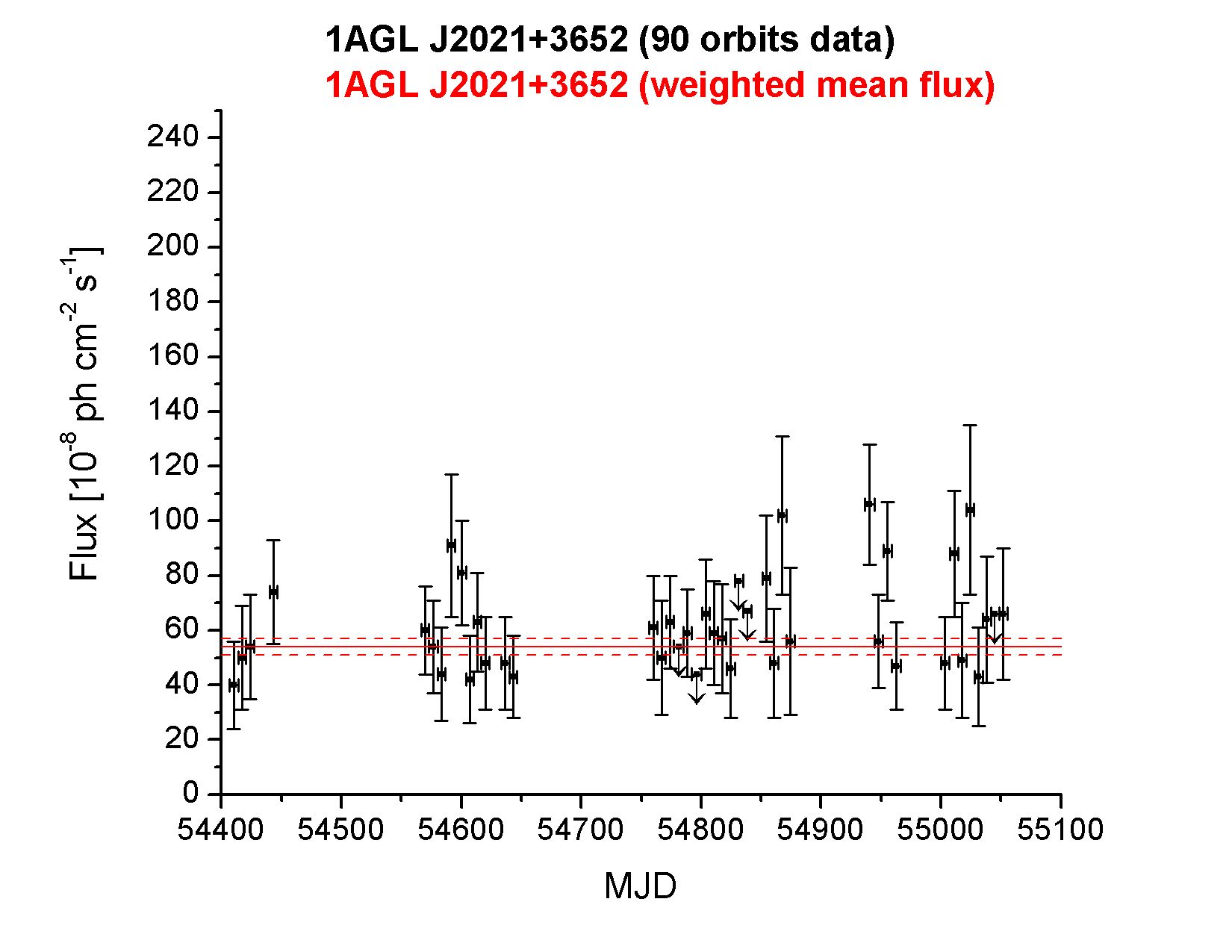}
	\end{tabular}
   }
   \caption{Flux for $E \ge 100~$MeV on six-day time intervals from 
	1AGL~J2022+4032 (\textit{left}) and 1AGL~J2021+3652 (\textit{right}). 
	The \textit{Red} line indicates the weighted mean of the 42 
	six-day fluxes, from which the $\chi^2$ calculated was calculated: 
	$(118\pm4) \times 10^{-8}\mathrm{~photons~cm}^{-2}\mathrm{~s}^{-1}$ 
	for 1AGL~J2022+4032, and 
	$(54\pm3) \times 10^{-8}\mathrm{~photons~cm}^{-2}\mathrm{~s}^{-1}$ for 
	1AGL~J2021+3652.} 
   \label{flux-graph-100MeV}
\end{figure*}

Next, we divided the observations from November 2007 to August 2009 into 42 
discrete fixed-length time intervals of $\approx6$~days ($\approx90$~orbits) 
each, analyzing the \textit{AGILE} $\gamma$-ray flux from the position of 
1AGL~J2022+4032, $(l, b)=(78.23,~2.12)$, for $E \ge 100~$MeV and 
$E \ge 400~$MeV, while keeping its position fixed and all nearby sources fixed 
in flux and position. We performed the same analysis on the nearby 
$\gamma$-ray source 1AGL~J2021+3652, $3.5^{\circ}$ away from 1AGL~J2022+4032, 
which \citet{halpern} identified as \object{PSR J2021+365}, 
in order to account for the effects of systematic errors. 
Figure~\ref{flux-graph-100MeV} shows the light curves of the two sources.

\begin{figure*}[htbp]
   \centering
   \resizebox{\hsize}{!}{
	\begin{tabular}{cc}
	\includegraphics{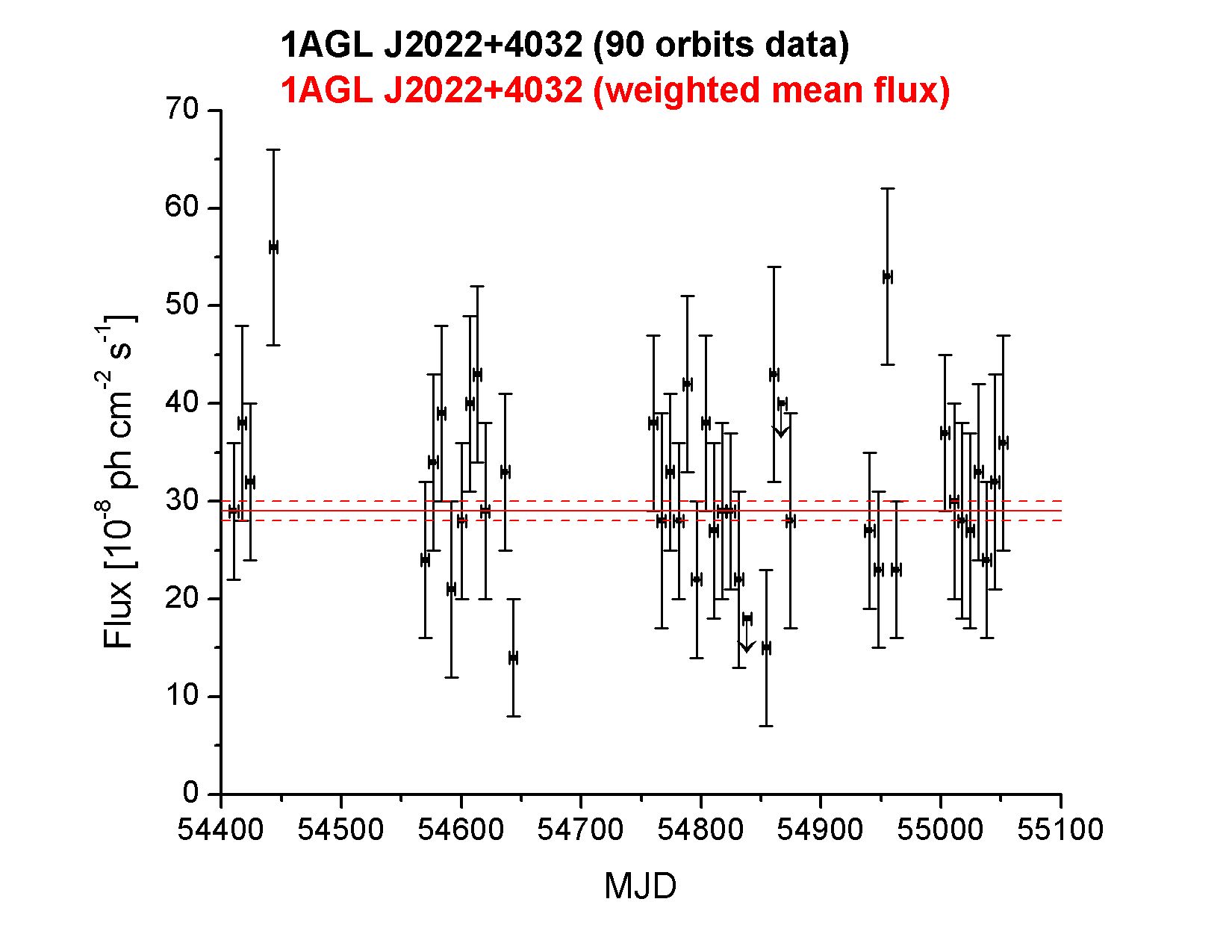} & 
	\includegraphics{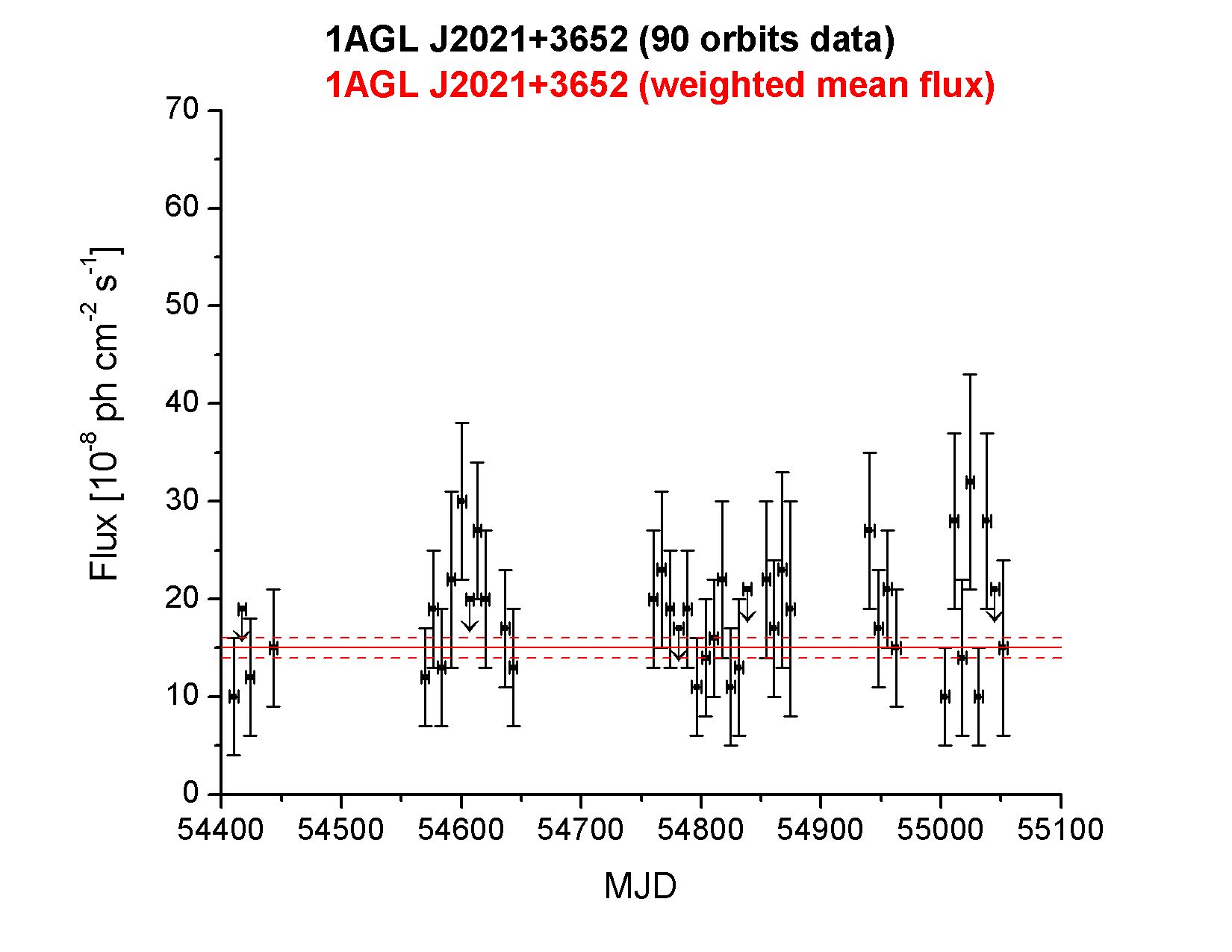}
	\end{tabular}
	}
   \caption{Flux for $E \ge 400~$MeV on six-day time intervals from 
	1AGL~J2022+4032 (\textit{left}) and 1AGL~J2021+3652 
	(\textit{right}).The \textit{Red} line indicates the weighted mean of 
	the 42 individual fluxes, from which the $\chi^2$ was calculated:
	$(29 \pm 1) \times 10^{-8}\mathrm{~photons~cm}^{-2}\mathrm{~s}^{-1}$ 
	for 1AGL~J2022+4032, and 
	$(15 \pm 1) \times10^{-8}\mathrm{~photons~cm}^{-2}\mathrm{~s}^{-1}$ 
	for 1AGL~J2021+3652.}
   \label{flux-graph-400MeV}
\end{figure*}

For $E \ge 400~$MeV, only the emission from the three steady sources shown in 
the lower half of Table~\ref{sources} were considered; J2033+4050 is not 
significantly detected for $E \ge 400~$MeV. In Figure~\ref{flux-graph-400MeV} 
we compare the light curves of 1AGL~J2022+4032 and 1AGL~J2021+3652.


\section{Discussion}

\subsection{Variability}

\begin{table*}[htbp]
   \centering
   \caption{Variability analysis with and without 10\% systematic errors on 
	fluxes.}
   \label{V100400}
   \begin{tabular}{c c c c c c c c}
\hline
Name            &   Syst. errors &  $\chi^2 (N_{df}=41)$  &       $P_{var}$      &   $V$    & $V_F$ & $P_{V_F}$ & $\delta F/F$\\
\hline\hline
1AGL J2022+4032, $E \ge 100~$MeV           &      yes    &    66.84   &       99.34\%        &  2.18  & 63.26 & 0.014 & 0.20\\
\hline
                                 &      no     &    82.45   &       99.99\%        &  3.88 & \multicolumn{3}{c|}{} \\
\hline \hline
1AGL J2021+3652, $E \ge 100~$MeV             &      yes    &    40.28   &       49.76\%        &  0.30  & 40.03 & 0.51 & 0.11\\
\hline
                                 &      no     &    44.47   &       67.24\%        &  0.48  & \multicolumn{3}{c|}{}\\
\hline
\hline\hline
1AGL J2022+4032, $E \ge 400~$MeV           &      yes    &    48.10   &       79.27\%        &  0.68 & 46.07 & 0.27 & 0.10 \\
\hline
                                 &      no     &    54.46   &       92.23\%        &  1.11  & \multicolumn{3}{c|}{}\\
\hline \hline
1AGL J2021+3652, $E \ge 400~$MeV             &      yes    &    34.05   &       22.95\%        &  0.11 & 32.77 & 0.82 & --- \\
\hline
                                 &      no     &    36.50   &       32.94\%        &  0.17  & \multicolumn{3}{c|}{}\\
\hline
   \end{tabular}
\end{table*}

We used the method developed by \citet{mclaughlin} to test the $\gamma$-ray 
flux variability of \object{1AGL J2022+4032} with respect to 
\object{1AGL J2021+3652}. A similar analysis for all the 1AGL sources is in 
preparation \citep{verrecchia}. The weighted mean flux is calculated from the 
fluxes in each 6-day time interval and their corresponding errors, from which 
the $\chi^2$ is derived. $Q$ is the probability that an intrinsically 
non-variable source (i.e. with constant flux) would produce by random chance a 
measured value of $\chi^2$ greater than or equal to the $\chi^2$ observed, and 
the variability index $V$ is defined as $V = - \log{Q}$. A source can be 
classified as \textit{nonvariable} if $V < 0.5$, \textit{uncertain} if 
$0.5 \le V < 1$, or \textit{variable} if $V \ge 1$. The value $V=1$ 
corresponds to a probability of variability $P_{var}=1-Q$ of 90\%.
Table~\ref{V100400} shows the value of $V$ for 1AGL~J2022+4032 and 
1AGL~J2021+3652 for $E \ge 100~$MeV and $E \ge 400~$MeV. For 1AGL~J2022+4032 
we find $V=2.18$ when systematic effects are included ($V=3.88$ for 
statistical only). For 1AGL~J2021+3652 the corresponding values are 
$V=0.30$ ($V=0.48$).

As a cross-check, we also calculated a complementary variability index, $V_F$, 
according to the formula used in the \textit{Fermi} catalogs 
\citep{fermibsl,1FGL}. This index is a simple $\chi^2$ where the weights 
include the systematic error, $f_{rel}$ which in our case is $10\%$, and the 
number of degrees of freedom is 41. The values, associated probabilities, and 
fractional excess variabilities above systematic and statistical fluctuations 
are found in Table~\ref{V100400}.

We find evidence for variability for $E \ge 100~$MeV in the emission from 
1AGL~J2022+4032 even allowing for systematic errors on the level of 10\%.  Any 
systematic effects that would influence the measurement of the flux should 
also have affected the nearby source 1AGL~J2021+3652, for which no 
corresponding variability is found.  However, 1AGL~J2021+3652 is only half as 
bright as 1AGL J2022+4032. Similarly, although we found no evidence for 
variability in the flux of 1AGL J2022+4032 for $E \ge 400~$MeV when systematic 
errors are taken into account,the average flux is only a quarter that of 
$E \ge 100~$MeV.  In both cases, the same intrinsic variability might be 
rendered undetectable because of reduced photon statistics.

\begin{figure}[htbp]
   \centering
   \resizebox{\hsize}{!}{
	\includegraphics{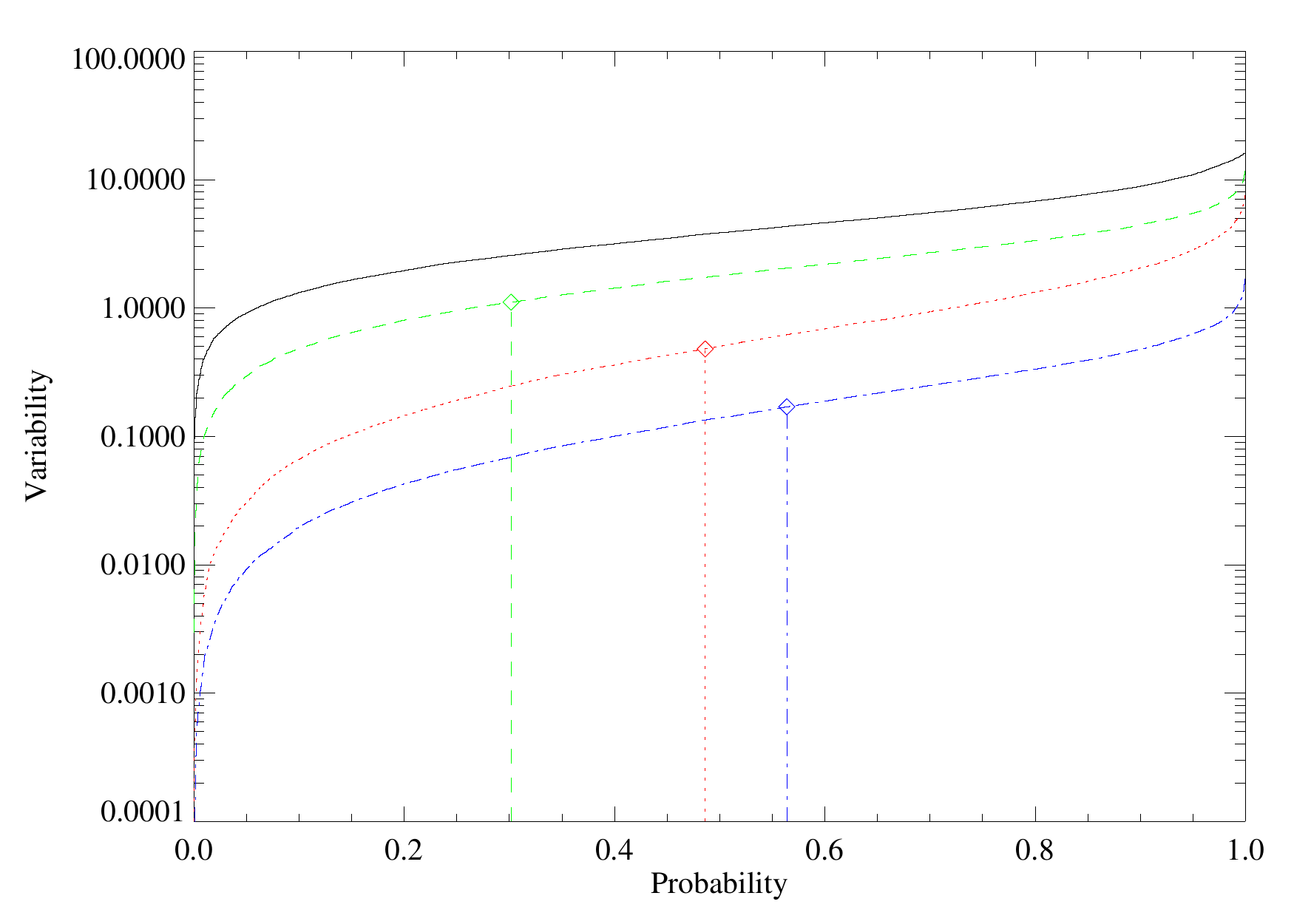}
	}
   \caption{Simulated cumulative distributions (10000 trials) of the 
	variability index $V$ versus fraction of trials, or probability. 
	{\it Solid black:} mean flux
	$131\times10^{-8}\mathrm{~photons~cm}^{-2}\mathrm{~s}^{-1}$ 
	and square root of variance 
	$33.8\times10^{-8}\mathrm{~photons~cm}^{-2}\mathrm{~s}^{-1}$ for 
	$E \ge 100~$MeV; {\it dotted red:} mean flux 
	$60\times10^{-8}\mathrm{~photons~cm}^{-2}\mathrm{~s}^{-1}$ 
	and square root of variance 
	$15.5\times10^{-8}\mathrm{~photons~cm}^{-2}\mathrm{~s}^{-1}$ for 
	$E \ge 100~$MeV; {\it dashed green:} mean flux
	$33\times10^{-8}\mathrm{~photons~cm}^{-2}\mathrm{~s}^{-1}$ 
	and square root of variance 
	$8.5\times10^{-8}\mathrm{~photons~cm}^{-2}\mathrm{~s}^{-1}$ for 
	$E \ge 400~$MeV; {\it dot-dahsed blue:} mean flux 
	$17\times10^{-8}\mathrm{~photons~cm}^{-2}\mathrm{~s}^{-1}$ 
	and square root of variance 
	$4.4\times10^{-8}\mathrm{~photons~cm}^{-2}\mathrm{~s}^{-1}$ for 
	$E \ge 400~$MeV. The vertical lines indicate the observed values of 
	$V$ and the associated probability for the corresponding dataset 
	(Table \ref{V100400}).}.
   \label{cumul}
\end{figure}

To determine whether the variability of 1AGL~J2022+4032 for $E \ge 100~$MeV 
would have been detectable if the source were half as bright (Case 1) or at 
higher energies (Case 2), we produced simulated observations of hypothetical 
sources with the same intrinsic variability, where "intrinsic" includes the
effects of systematic errors.  The meaning of "same" is somewhat ill-defined, 
since many qualitatively and quantitatively different parent distributions can 
produce the same variability index, or even the same observed fluxes within 
errors.  In order to simplify the relationship with the calculation of 
$\chi^2$, we assume that the intrinsic flux in each observation is drawn from 
a Gaussian distribution whose mean is equal to the mean flux of the source in 
order to find the variance which reproduces the observed value of 
$V=3.88$ for 1AGL~J2022+4032. We determined that, given the exposures of each 
of the 42 time intervals, an intrinsic variability of 26\% (square root of 
variance $33.8\times10^{-8}\mathrm{~photons~cm}^{-2}\mathrm{~s}^{-1}$ for 
$131\times10^{-8}\mathrm{~photons~cm}^{-2}\mathrm{~s}^{-1}$, the mean flux of 
1AGL~J2022+4032 for $E \ge 100~$MeV) produces a distribution in $V$ with a 
median equal to the observed value, 3.88 (Figure~\ref{cumul}, solid black).

For Case 1, we simulated 10000 series of 42 observations for a source with 
intrinsic fluxes taken from a Gaussian distribution with mean equal to 
$60\times10^{-8}\mathrm{~photons~cm}^{-2}\mathrm{~s}^{-1}$ 
and square root of variance of 
$15.5\times10^{-8}\mathrm{~photons~cm}^{-2}\mathrm{~s}^{-1}$ 
using the same exposures as \object{1AGL J2022+4032} for $E \ge 100~$MeV 
(Figure~\ref{cumul}, dotted red).  The median value of $V$ is 0.50.  
$V \leq 0.48$ is produced in $49\%$ of trials, while $V \ge 1.0$ in $28\%$ of 
trials.  The same level of intrinsic variability would be likely to produce a 
variability index similar to that of 1AGL~J2021+3652, but would not be likely 
to be classified as variable, if the source were half as bright. For Case 2, 
we simulated 10000 series of 42 observations for a source with intrinsic 
fluxes taken from a Gaussian distribution with mean equal to 
$33\times10^{-8}\mathrm{~photons~cm}^{-2}\mathrm{~s}^{-1}$ 
and square root of variance 
$8.4\times10^{-8}\mathrm{~photons~cm}^{-2}\mathrm{~s}^{-1}$, 
using the exposures of 1AGL~J2022+4032 for $E \ge 400~$MeV 
(Figure~\ref{cumul}, dashed green).  The median value of $V$ is 1.75. 
$V \leq 1.11$ is produced in $31\%$ of trials, while $V \ge 1.0$ in 
$74\%$ of trials.  The same level of flux variability at high energies would 
be detected more often than not. Equivalent intrinsic variability at the 
flux level of 1AGL~J2021+3652 for $E \ge 400~$MeV would be completely 
undetectable (Figure~\ref{cumul}, dot-dashed blue).

\subsection{Analysis of pulsed and unpulsed components}

\begin{figure}[htbp]
   \centering
   \resizebox{\hsize}{!}{
	\includegraphics{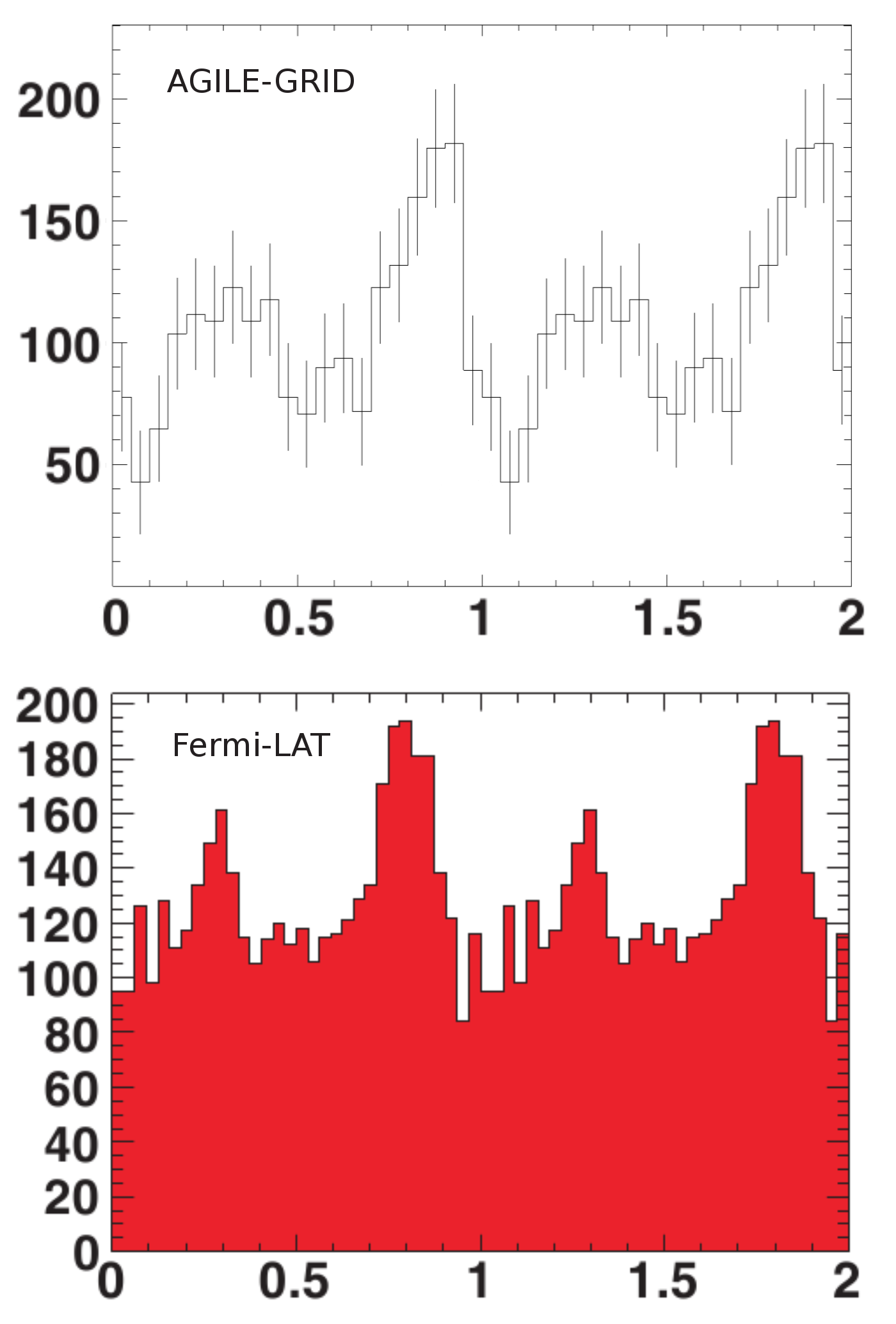}
	}
   \caption{$\gamma$-ray phase profile of PSR J2021+4026 as seen by
	\textit{AGILE} (\textit{top}) and \textit{Fermi} (\textit{bottom}). 
	The \textit{AGILE} profile is for photons with $E \ge 100$~MeV while 
	the \textit{Fermi} profile, taken from \citet{fermipulsar}, is for 
	$E \ge 300$~ MeV.}
   \label{pulsarlc}
\end{figure}

We attempted to perform an analysis of the on- and off-peak components of the 
1AGL~J2022+4032 \textit{AGILE} data with respect to the \textit{Fermi} 
$\gamma$-ray ephemeris of LAT~\object{PSR J2021+4026} in order to test whether 
the apparent flux variability is due to a source other than the pulsar. 
However, as shown in Table 3 of the \textit{Fermi} Pulsar Catalog 
\citep{fermipulsarcatalog}, LAT~PSR J2021+4026 is one of only two 
\textit{Fermi} pulsars for which the $\gamma$-ray phase profile is difficult 
to separate into on-peak and off-peak phases, both because the $\gamma$-ray 
peaks of the pulsar are very broad and because the $\gamma$-ray emission has a 
high unpulsed fraction (Figure~\ref{pulsarlc}). We attempted a number of cuts 
on the \textit{AGILE} data. Varying the off-peak exposure fraction from 
10\% to 45\% yielded a monotonic variation in flux with no obvious plateau. 
The counts maps show evidence of the pulsar contribution for all but the 
lowest off-peak exposure fraction, for which the statistical significance 
$\sqrt{TS} = 8.4$ over two years of data is insufficient to perform a 
variability analysis.  For these reasons, a credible analysis of the on-peak 
and off-peak variability based on the \textit{AGILE} data was not feasible.

\section{Discussion of possible counterparts}

\begin{figure*}[htbp]
\sidecaption
   \includegraphics[width=12cm]{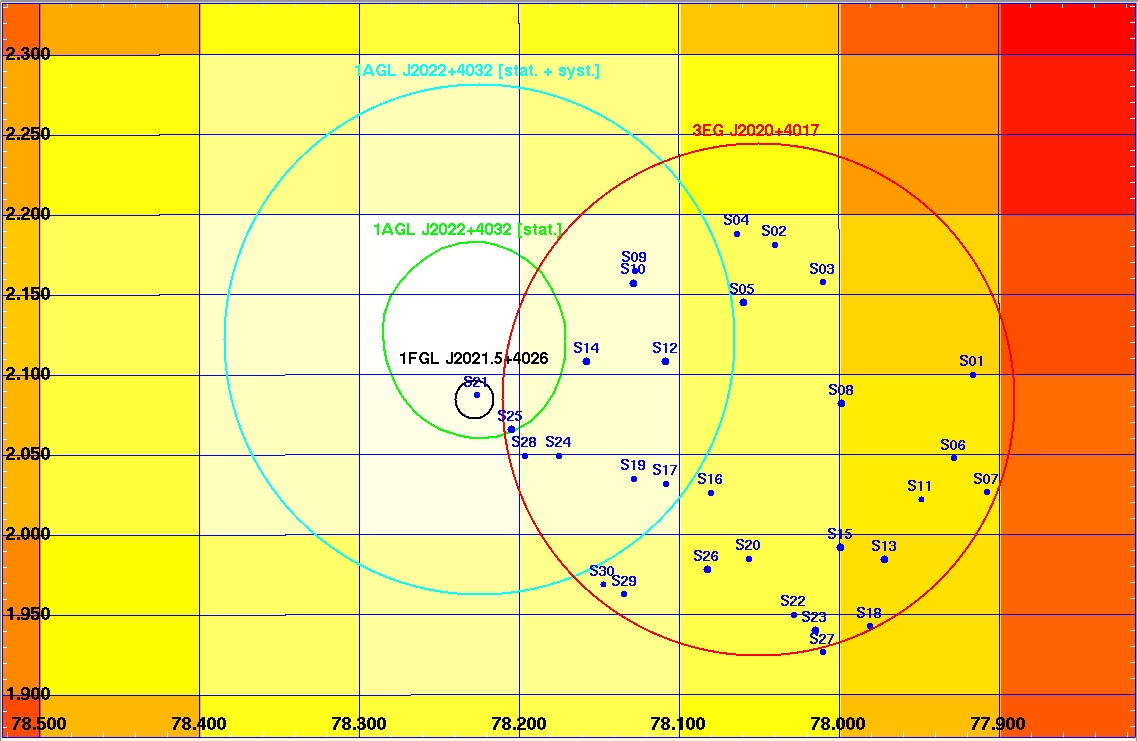}
   \caption{Possible x-ray counterparts from \citet{weisskopf}. \textit{Green} 
	contour level: \textit{AGILE-GRID} 95\% C.L. (only statistical) for 
	1AGL~J2022+4032; \textit{cyan} circle: \textit{AGILE-GRID} errorbox 
	(statistical + systematic error); \textit{red} circle: \textit{EGRET} 
	errorbox for 3EG~J2020+4017; \textit{black} circle: \textit{Fermi} 
	errorbox for 1FGL~J2021.5+4026.}
   \label{Chandra}
\end{figure*}

In Figure \ref{Chandra}, we show the X-ray sources listed in
\citet{weisskopf} as possible counterparts for \object{3EG J2020+4017}.
The \textit{AGILE} contour of the persistent source is consistent with
the position of the X-ray source \object{[WSC2006] S21}, which has been
associated with the LAT~\object{PSR J2021+4026} whose pulsations were 
discovered by \textit{Fermi}. \citet{trepl} searched the {\it XMM-Newton} 
archival data and found 2XMM~J202131.0+402645, a point source coincident with 
S21. However, re-analyzing the {\it Chandra} data, they found that 
\object{[WSC2006] S25}, a strong point source within the \textit{Fermi} 0FGL 
error box although outside the \textit{Fermi} 1FGL error box, showed evidence 
of variability during the {\it Chandra} observation. In addition, S25 is not 
visible in the {\it XMM} data, indicating long-term X-ray variability. If S25 
is also a variable $\gamma$-ray source, it would be within the \textit{AGILE} 
source location accuracy of $\sim 1^{\circ}$ for the 1-2 week observation 
durations, and could be responsible for the variability apparently observed by 
\textit{AGILE} below 400 MeV. However, S25 has an infrared counterpart and 
could be a normal star.

The \textit{Fermi} 11-month source catalog \citep{1FGL} detected a nearby 
source, 1FGL~J2020.0+4049, associated with TeV source VER~J2019+407, that is 
within the one-week error circle of \textit{AGILE}. It was not detected by 
\textit{AGILE}, probably because of its hard spectrum with index 
$-2.12\pm 0.08$. Its \textit{Fermi} light curve is consistent with no 
variability at a level of 33\%; however, its measured flux for $E>100$~MeV 
decreased from 
$(22\pm6)\times10^{-8}\mathrm{~photons~cm}^{-2}\mathrm{~s}^{-1}$ in the first 
month to below $4.3\times10^{-8}\mathrm{~photons~cm}^{-2}\mathrm{~s}^{-1}$ by 
the seventh month.  An intrinsic flux variability of 
$34\times10^{-8}\mathrm{~photons~cm}^{-2}\mathrm{~s}^{-1}$ over the longer 
\textit{AGILE} observation period would explain the apparent flux variability 
of 1AGL~J2022+4032.

Detection of simultaneous flaring in $\gamma$-ray and, e.g., X-rays would 
provide a definitive identification. However, soft X-ray instruments have 
fields of view too narrow to make triggered observations practical. We note 
that a special class of microquasars may not necessarily produce simultaneous 
hard X-ray emission \citep[][hereafter RV09]{romero}.

\subsection{A variable pulsar}

It is possible that the pulsar LAT~\object{PSR J2021+4026} itself has variable 
$\gamma$-ray emission for the energy range below 400~MeV. This scenario, if 
confirmed, would open a new field of investigation of $\gamma$-ray pulsars. 
However, in the absence of other information supporting this hypothesis pulsar 
$\gamma$-ray variability of LAT~PSR J2021+4026 is unlikely. This conclusion is 
based on both previous observational evidence from \textit{EGRET} and from 
general theoretical considerations. In particular, the spin period and time 
derivative \citep{fermipulsar} of LAT~PSR J2021+4026 
($P$ = 265 ms, $\dot{P}$ = $54.8 \times 10^{-15}$) assign a unremarkable 
position to this pulsar in both the  $P-\dot{P}$ diagram and 
in the pulsar $\gamma$-ray luminosity vs. period (or Goldreich-Julian current 
diagram). Instabilities in the $\gamma$-ray pulsed emission of 
LAT~PSR~J2021+4026 might be associated with radio and/or X-ray signal changes 
during the period of the detected $\gamma$-ray variability. Future monitoring 
of LAT~PSR~J2021+4026 can contribute to test this fascinating hypothesis.

\subsection{Background blazar}

Another possibility is that a blazar behind the Galactic plane is contributing 
variable $\gamma$-ray emission to the measured flux of 1AGL~J2022+4032.  There 
are a number of known blazars in the Cygnus region, but they are well outside 
the typical \textit{AGILE} one-week error circle. Most optical and radio AGN 
catalogs avoid the Galactic plane because of the high concentration of 
Galactic sources, heavy extinction, and/or diffuse radio emission 
\citep{fermiagn}. Each of these difficulties is especially acute at the
position of 1AGL~J2022+4032, which is located within the Gamma-Cygni supernova 
remnant shell (see Fig.~\ref{gamma_cyg_snr}). \citet{trepl} found multiple 
areas of concentrated radio emission within the \textit{Fermi} error box of 
\object{1FGL J2021.5+4026} as well as evidence of variable X-ray emission from 
nearby source \object{[WSC2006] S25}. In addition, the $\gamma$-ray flux could 
be well below the detection threshold of \textit{AGILE} (which, in the 
presence of a known, bright $\gamma$-ray source is quite high) and still 
contribute to the overall $\gamma$-ray variability within the one-week 
\textit{AGILE} error box. Knowing that the intrinsic distribution of blazars 
should be isotropic, we can use the AGN associations above $|b|>10^{\circ}$ in 
the First AGN Catalog \citep[][hereafter 1LAC]{fermiagn} to estimate the 
probability of finding at least one blazar within the $\approx 1^{\circ}$ 
error box of a one-week observation with \textit{AGILE}.  The probability of 
finding a blazar similar to the 599 1LAC associations is $\approx 0.05$. A 
more conservative estimate using only the 281 1LAC associated FSRQs, yielding 
a probability of $\approx 0.02$ of finding at least one FSRQ, would better 
represent the need for $\gamma$-ray variability and the non-detection by 
\textit{Fermi}. In either case, the probability of chance coincidence is quite 
low.

\subsection{An X-ray quiet microquasar}

Taking into account both the \textit{AGILE-GRID} emission above 100~MeV and 
the {\it Super-AGILE} upper limit in the 15-60 keV range ($\sim60$ mCrab), we 
consider  the possibility that the detected $\gamma$-ray variability is caused 
by transient activity of an X-ray quiet microquasar. RV09 analyzed several 
Galactic sources with variable emission in the $\gamma$-ray energy range and 
showing a ratio $L_{\gamma}/L_{X} \gg 1$. They proposed that this kind of 
emission (shortly variable, X-ray quiet, \dots) can be produced by 
proton-dominated jets in a special class of Galactic microquasars. The bulk of 
the emission at $\gamma$-ray  energies is  produced by hadronic jets emitted 
from an accreting source. The model of RV09 predicts a $\gamma$-ray luminosity 
for this process on the order of $\approx 10^{34}$ $\mathrm{erg/s}$. Assuming 
the presence of a $\gamma$-ray source (an X-ray quiet microquasar) within 
the error box of LAT~PSR~J2021+4026, we find that that it is required to be at 
a distance of $\approx300$~pc from the Earth, i.e., closer than the pulsar 
(1-2 kpc). 

The probability of finding this particular type of microquasar within the 
error box of 1AGL~J2022+4032 is difficult to quantify.  Nevertheless, because 
X-ray binaries are concentrated in the star-forming regions in the Galactic
plane, and high-mass X-ray binaries particularly along tangents of spiral arms
such as the Cygnus region \citep{liu07,liu06}, the likelihood that there is 
an appropriate microquasar within the error box is much higher than that of 
blazars, which are isotropically distributed. Similar reasoning applies to 
such possible source types such as massive stellar winds \citep{etacar} and 
novae \citep{v401cyg}.


\section{Conclusions}

The apparent $\gamma$-ray flux variability of \object{1AGL J2022+4032} in the 
100-400~MeV range as seen by \textit{AGILE} leads us to consider two possible 
explanations. One is that the \textit{Fermi} pulsar, 
LAT~\object{PSR J2021+4026} itself has a variable $\gamma$-ray flux. This 
behavior would be highly unusual given the properties of $\gamma$-ray (and 
radio) pulsars with similar characteristics. Based on the large $\gamma$-ray 
unpulsed fraction of LAT~PSR~J2021+4026 as seen by its folded light curve 
\citep{fermipulsar}, we believe that it is more likely that another variable 
$\gamma$-ray source within the error box of LAT~PSR~J2021+4026 contributes to 
the total emission of 1AGL~J2022+4032. 1AGL~J2022+4032 was not detected in the 
hard X-rays during the \textit{AGILE} observation periods, with upper limits 
in {\it Super-AGILE} varying from 10 to 50-60~mCrab in the 15-60~keV range. 
Therefore, if the variable source is a microquasar rather than an unidentified
low-frequency blazar, it would be an X-ray-quiet $\gamma$-ray variety, 
possibly of the type proposed by RV09.

We note that during the period in which 1AGL~J2022+4032 was observed by
\textit{Fermi} (starting in August 2008), its flux in \textit{AGILE-GRID} was 
slightly more stable.  In fact, using only the \textit{AGILE} fluxes for 
$E \ge 100~$MeV from the 28 time intervals during the \textit{Fermi} 
observations, we find that the variability $V$ was $1.65$ without systematic 
errors and $0.96$ with systematic errors, while in the period before 
\textit{Fermi} the variability $V$ was $1.94$ without systematic errors and 
$1.23$ with systematic errors. Future observations of 1AGL~J2022+4032 by both 
\textit{AGILE} and \textit{Fermi} will reveal whether this fascinating source 
continues to show evidence over the long term.

\end{document}